\documentclass[aps,prl,twocolumn,showpacs,superscriptaddress]{revtex4-1}

\usepackage[utf8]{inputenc}
\usepackage{graphicx}  
\usepackage{epstopdf,epsfig}
\usepackage{amsmath}
\usepackage{amssymb}
\usepackage{amsbsy,amsxtra}
\usepackage{amsfonts}
\usepackage{braket}
\usepackage{dcolumn}   
\usepackage{bm}        
\usepackage{amssymb}   
\usepackage[usenames]{color}
\usepackage{natbib}
\usepackage{float}
\usepackage{gensymb}
\usepackage{subfigure}
\usepackage{lipsum}

\begin{document}

\title{Electron degeneracy and intrinsic magnetic properties of epitaxial Nb:SrTiO$_3$ thin-films controlled by defects}

\author{A. Sarantopoulos}
\affiliation{Centro de Investigaci\'on en Qu\'imica Biol\'ogica y Materiales Moleculares (CIQUS), Universidad de Santiago de Compostela, 15782-Santiago de Compostela, Spain}
\author{E. Ferreiro-Vila}
\affiliation{Centro de Investigaci\'on en Qu\'imica Biol\'ogica y Materiales Moleculares (CIQUS), Universidad de Santiago de Compostela, 15782-Santiago de Compostela, Spain}
\author{V. Pardo}
\affiliation {Departamento de F\'isica Aplicada, Universidade de Santiago de Compostela, Spain}
\affiliation{Instituto de Investigaci\'ons Tecnol\'oxicas, Universidade de Santiago de Compostela, Santiago de Compostela, Spain}
\author{C.Magen}
\affiliation{Laboratorio de Microscopías Avanzadas, Instituto de Nanociencia de Aragón, Universidad de Zaragoza, 50018 Zaragoza, Spain}
\affiliation{Departamento de Física de la Materia Condensada, Universidad de Zaragoza, 5009 Zaragoza, Spain}
\affiliation{Fundaci\'on ARAID, 50018 Zaragoza, Spain}
\author{M. H. Aguirre}
\affiliation{Laboratorio de Microscopías Avanzadas, Instituto de Nanociencia de Aragón, Universidad de Zaragoza, 50018 Zaragoza, Spain}
\affiliation{Departamento de Física de la Materia Condensada, Universidad de Zaragoza, 5009 Zaragoza, Spain}
\author{F. Rivadulla}
\email{f.rivadulla@usc.es}
\affiliation{Centro de Investigaci\'on en Qu\'imica Biol\'ogica y Materiales Moleculares (CIQUS), Universidad de Santiago de Compostela, 15782-Santiago de Compostela, Spain}


\begin{abstract}

We report thermoelectric power experiments in e-doped thin films of SrTiO$_3$ (STO) which demonstrate that the electronic band degeneracy can be lifted through defect management during growth. We show that even small amounts of cationic vacancies, combined with epitaxial stress, produce a homogeneous tetragonal distortion of the films, resulting in a Kondo-like resistance upturn at low temperature, large anisotropic magnetoresistance, and non-linear Hall effect. Ab-initio calculations  confirm a different occupation of each band depending on the degree of tetragonal distortion. The phenomenology reported in this paper for tetragonally distorted e-doped STO thin films, is similarto that observed in LaAlO$_3$/STO interfaces and magnetic STO quantum wells.

\end{abstract}

\maketitle


The possibility of growing epitaxial interfaces with atomic precision leads to unexpected functionalities in polar interfaces between oxide insulators \cite{Nakagawa, Hwang_NatMat, CBE_Science}.
In particular, the two-dimensional electron gas (2DEG) emerging at the interface between SrTiO$_3$ (STO) and LaAlO$_3$ (LAO), \cite{Ohtomo, Sing} can be tuned to produce a very rich phase diagram, including magnetism and superconductivity.\cite{Brinkman2007, Reyren} These properties are common to the 2DEGs stabilized at the bare surface of bulk STO, and are derived to a large extent from the arrangement and filling of the Ti-$t_{2g}$-derived sub-bands close to the conduction band minimum.\cite{MacDonald2011,Santander-Syro2010,Meevasana2011,King2014}
This orbital reconstruction can be intrinsic, to cancel the electrostatic energy of the polar interface, \cite{Salluzzo} but could also be caused by the structural distortions and atomic vacancies which relax the epitaxial strain in these thin-film heterostructures.
However, the effect of cationic and anionic vacancies on the transport properties of both STO \cite{Siemons2007} and LAO/STO interfaces \cite{Breckenfeld2013} is mostly considered from the point of view of their acceptor/donor nature over the total charge density, although vacancies distribute along the crystal structure expanding the unit cell due to an increased Coulomb repulsion.\cite{Keeble2013} In the case of epitaxial thin films below a certain critical thickness, this has to be necessarily accomodated purely as an increase of the c-axis parameter, due to the clamping of the in-plane lattice parameters to the substrate.\cite{BreckenfeldJMCC} In this case, if the tetragonal distortion is homogeneous throughout the film instead of localized around defects, an important effect over the band structure can be anticipated. 

Here we show that the formation of cation/anion vacancies during epitaxial growth results in a homogeneous tetragonal distortion along the films, which determines their transport properties. Ab initio calculations and thermoelectric power experiments under different degrees of stress suggest that the band degeneracy characteristic of STO can be lifted for sufficiently distorted films. A resistance upturn at low temperature, large anisotropic magnetoresistance (AMR) and non linear Hall effect are observed. This phenomenology was previously observed in LAO/STO interfaces and magnetic STO quantum wells.\cite{Budhani2014,Stemmer2014} Here we demonstrate that it can also be observed in electron-doped STO thin films under a sufficiently large tetragonal distortion.


Epitaxial thin films of Nb doped SrTiO$_3$ (Nb:STO) were grown by PLD (KrF, $\lambda = 248$ nm, laser fluence 0.9 J/cm$^{2}$, 5 Hz, at 800 \degree C) on top of (001) TiO$_{2}$-terminated STO (a = 3.905 {\AA}), and (001) (LaAlO$_{3}$)$_{0.3}$(Sr$_{2}$AlTaO$_{6}$)$_{0.7}$ (LSAT, a = 3.87{\AA}) substrates. The oxygen pressure was varied from 10$^{-2}$ to 200 mTorr. After  deposition the samples were cooled at the same oxygen pressure. The conditions were optimized to obtain layer-by-layer growth, even in films as thick as $\approx$ 50 nm (see supporting information).



\begin{figure}[htbp]
	\centering
		\hspace{-6px}\includegraphics[width=0.50\textwidth]{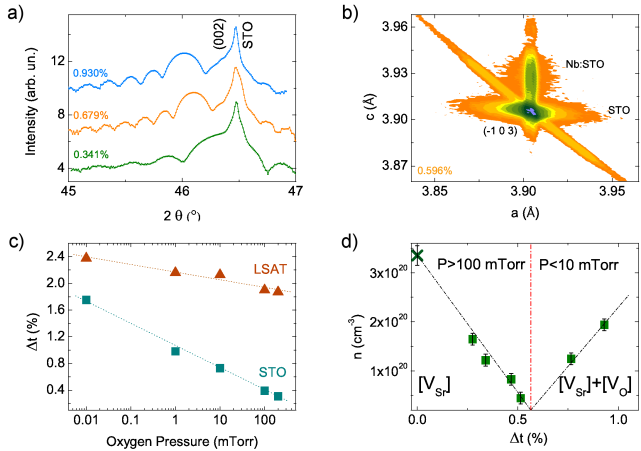}
	\vspace{-20px}\caption{\footnotesize{a) XRD patterns of 25(1) nm thick Nb:STO films grown on (001) STO, at different oxygen pressures. Numbers indicate the percentage of tetragonal distortion $\Delta$t. b) High resolution reciprocal space map around the (-103) reflection, for one film deposited at 1 mTorr. c) Pressure dependence of the tetragonal distortion for thin films deposited on STO and LSAT. d) Room temperature carrier density as a function of the tetragonal distortion, determined from Hall effect for samples deposited on STO.  The cross corresponds to the theoretical value expected for the nominal Nb-doping}}
	\label{Imagen1}
\end{figure}

\begin{figure}[htbp]
	\centering
		\hspace{-6px}\includegraphics[width=0.50\textwidth]{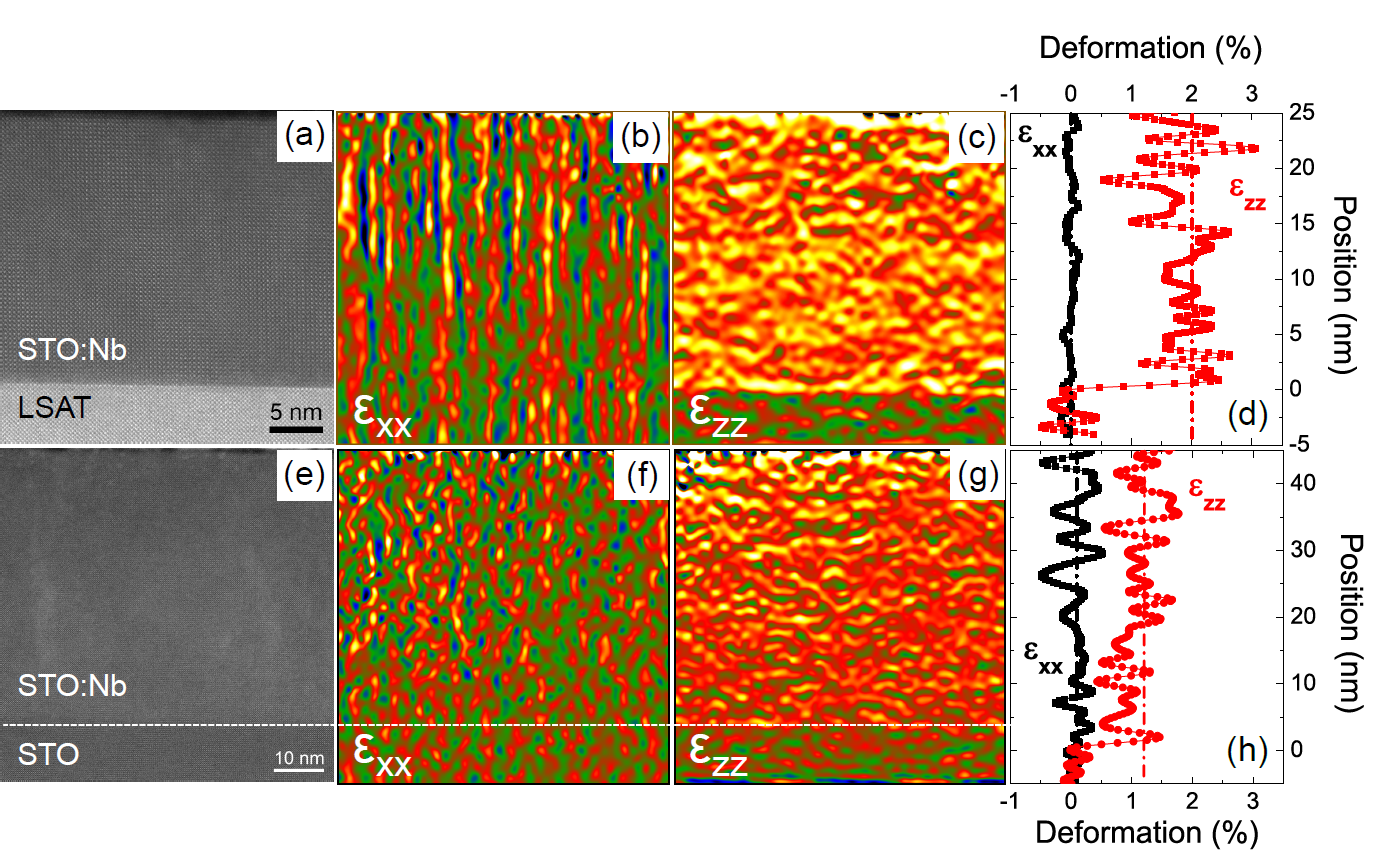}
	\vspace{-20px}\caption{\footnotesize{HAADF-STEM image of Nb:STO thin films deposited on STO (bottom) and LSAT (top). GPA analysis showing the in-plane $\varepsilon_{xx}$ and out of plane $\varepsilon_{zz}$ elongation, with respect to the substrate.  A profile along the film thickness is shown in panels d) and h). The analysis indicates a total tetragonal distortion $\Delta$t $\approx$1.5$\%$ for STO and $\approx$2.5$\%$ for LSAT.}}
	\label{Figure_TEM}
\end{figure}

Growing stoichiometric films of STO by PLD is only possible in a very narrow range of laser fluence, oxygen pressure and temperature.\cite{Ohnishi2005, Keeble2010, Kozuka2010, Wicklein2012, BreckenfeldChemMat, Keeble2013, BreckenfeldJMCC} Away from this window, the samples always present a varying amount of cationic and anionic vacancies, which produce a substantial enlargement of the unit cell.\cite{Ohnishi2005}
The effect of oxygen pressure on the microstructure of the films studied in this work is summarized in Figures \ref{Imagen1} and \ref{Figure_TEM}. As expected, reducing the pressure results in a substantial increase of the c-axis lattice parameter (Fig. \ref{Imagen1}a).\cite{Ramesh2011} However, our study adds some important findings to previous works: X-ray analysis shows that the epitaxial films grow fully strained at all pressures (Fig. \ref{Imagen1}b). High Angle Annular Dark Field- Scanning Transmission Electron Microscopy and Geometrical Phase Analysis (HAADF-STEM and GPA) show that the expansion of the c-axis parameter is homogeneous along the film thickness, even for 50 nm thick films (Fig. \ref{Figure_TEM}h), not influenced by the presence of clustered defects. Therefore, it can be concluded that there is a homogeneous tetragonal distortion of the unit cell as the oxygen pressure is decreased. This distortion can be quantified by $\Delta$t=(c$_f$-a$_s$)$\times$100/a$_s$, where $f$, and $s$ stand for film and substrate, respectively.


At first sight it is tempting to adscribe the distortion to the presence of oxygen vacancies, V$_{O}$. However, post-annealing of the films in oxygen resulted in a negligible variation of the lattice parameter (see supporting information), or even a slight increase.\cite{Ohnishi2010} Therefore, V$_{O}$ cannot be the main source of this expansion.
Other possible sources of expansion could be the formation of defects due to bombardment of the film with highly energetic ions during deposition\cite{JPMaria, DamodaranAdvMat}, or the presence of cationic vacancies.\cite{Ohnishi2010, Schlom2009, BreckenfeldChemMat, BreckenfeldJMCC} In the range of laser fluence and oxygen pressures used in this work, Sr vacancies (V$_{Sr}$) are energetically favored over V$_{Ti}$.\cite{Kozuka2010}
We performed XPS analysis to determine the stoichiometry of our samples, and confirmed an increasing amount of Sr vacancies as oxygen pressure decreases (see supporting information). Within the resolution of our analysis, the samples grown at 100-200 mTorr are nearly stoichiometric, and the amount of V$_{Sr}$ increases to $\approx$5-7$\%$ at  lower pressures. The analyses also confirm the equal amount of Nb ($\approx$1.8(1)$\%$) in all samples.
Therefore, although the presence of V$_{O}$ is expected to be increasingly important as the oxygen pressure decreases, we conclude that the major contribution to the lattice expansion in the whole pressure range is due to doubly charged V$_{Sr}$. It is very interesting that a small concentration of cationic vacancies  is able to produce a homogeneous tetragonal distortion of the whole film, as shown in Fig. \ref{Figure_TEM}.

The increasing influence of oxygen vacancies in the whole pressure range can be further studied by Hall effect, given the different nature of V$_{Sr}$ (e-acceptor),\cite{Tanaka} and V$_{O}$ (e-donor) over the total charge density.

The electronic density determined by Hall measurements at room temperature is shown in Figure \ref{Imagen1}d for a series of films deposited on STO. In agreement with the previous discussion, the charge density shows two distinct regimes as a function of $\Delta$t: for samples grown at high oxygen pressure the carrier density decreases continuously from the expected nominal value, reflecting the  presence of electron-trapping defects.
Besides, growing samples below $\approx$10 mTorr results in an increase of the carrier density, showing the important contribution of oxygen vacancies to the electronic transport below this pressure range.

Therefore, there is a continuous change in the nature of vacancies as a function of the oxygen pressure. However, while they show an opposite contribution to the resistivity, only V$_{Sr}$ produce a significant contribution to the tetragonal distortion of the lattice.


Also, the analysis of the structural and microstructural data shows a different volume for the films grown on STO and LSAT (Fig. \ref{Imagen1}c), demonstrating a dependence of the unit-cell expansion on the epitaxial stress. Although growing Nb:STO on LSAT introduces a moderately compressive stress ($\approx$0.9$\%$), the continuous increase of the c-axis parameter with decreasing oxygen pressure is totally unexpected, on the basis of Poisson's effect in a stoichiometric system.

Once the existence of a homogeneous tetragonal distotion was established, we studied its effect on the electronic band structure of the system. We have performed density functional theory-based calculations\cite{dft,dft_2} on bulk STO under various degrees of tetragonal distortion $\Delta$t$ \approx $0-3\% (allowing for octahedral rotations). 

Spin-orbit coupling splits the t$_{2g}$ states (effectively an l=-1 triplet)\cite{fazekas} at the $\Gamma$-point, into a low-lying doublet and an upper lying singlet (see Fig. 3a where the band structure for $\Delta t = 2 \%$ is shown). The additional tetragonal distortion splits the doublet, the so-called j$_{eff}$= 3/2 states, which in the cubic case are two Kramers doublets. In order to analyze the evolution of the occupation of the different bands as $\Delta t$ increases, we focused on the number of electrons with $d_{xy}$ symmetry ($n_{xy}$) as a function of the total carriers (this was done assuming all the carriers in the unit cell are t$_{2g}$ electrons ($xy + xz + yz$)).  
The on-site energy of the d$_{xy}$ electrons increases with $\Delta$t, as they are relatively destabilized by the octahedral elongation with respect to the d$_{xz,yz}$.
Therefore, a reduction of the relative weight of the $xy$ electrons is anticipated from the calculations as $\Delta t$ and $n$ increase, thus reducing the overall electron degeneracy inside the t$_{2g}$ manifold. This trend is observed in Fig. 3b, in the range of doping relevant for comparison with our experiments.


\begin{figure}[htbp]
	\centering
		\hspace{-6px}\includegraphics[width=0.50\textwidth]{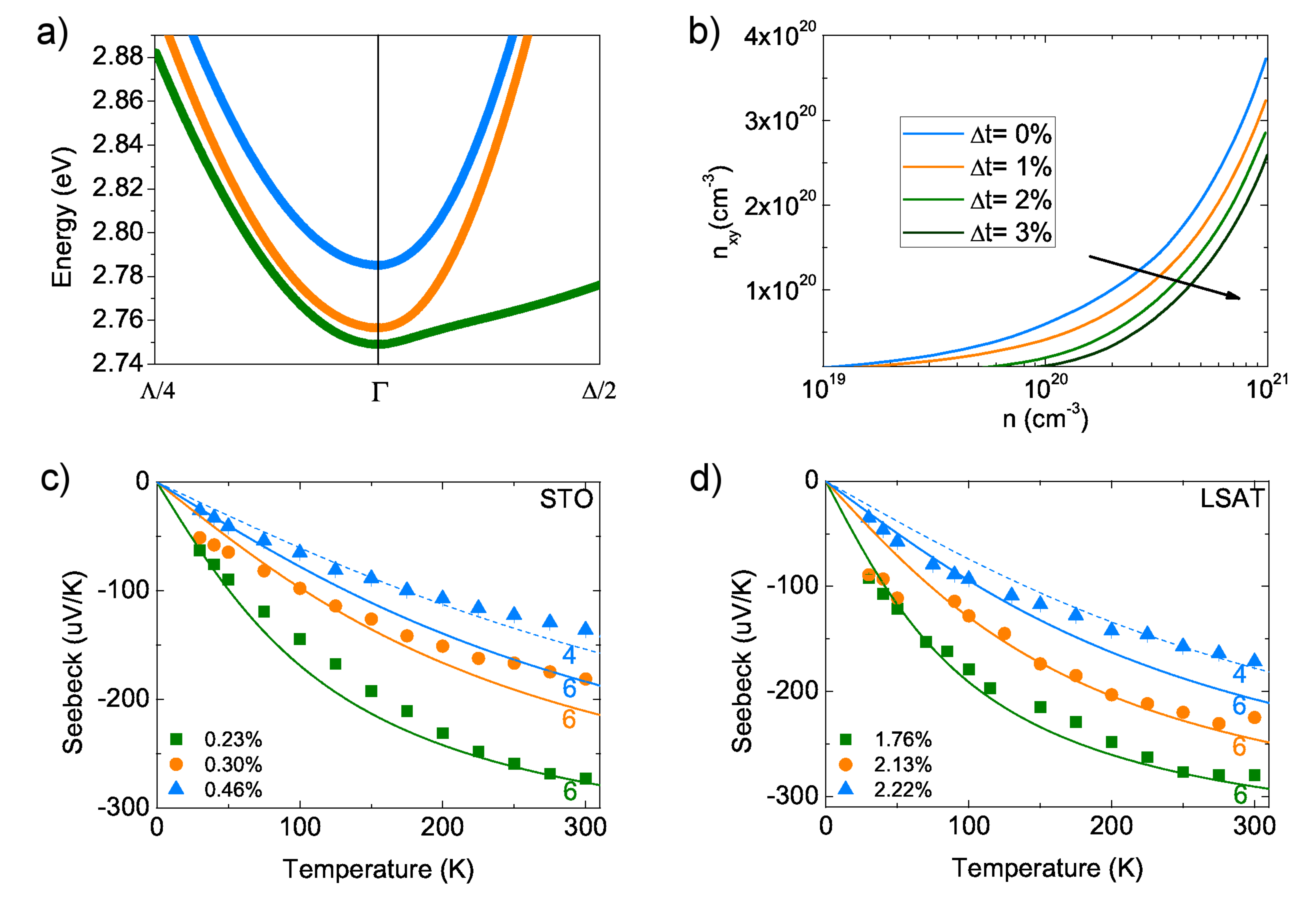}
	\vspace{-20px}\caption{\footnotesize{a) Band structure of bulk STO under a tetragonal distorsion $\Delta$t= 2\% with the upper band split by spin-orbit coupling. The low-lying doublet is split as a result of the tetragonal distortion. b) Total number of electrons with d$_{xy}$ symmetry as a function of the total number of $d$ electrons, which decreases as the tetragonal distortion increases. c) Temperature dependence of the Seebeck coefficient for thin films deposited on STO and d) LSAT, for different $\Delta$t. The lines are calculations from equation \ref{Seeb-gen}.}}
	\label{Imagen3}
\end{figure}


The impact of this change in the band structure and occupation can be tested experimentally from the temperature dependence of the Seebeck coefficient, for samples under different degree of distortion. Given the band structure of STO, a good approximation for the thermoelectric power of carriers at the bottom of its conduction band could be derived from the Boltzmann equation for a parabolic band ($z$-fold degenerate):\cite{Tokura2001}

\begin{equation}
\label{Seeb-gen}
S = \left( \frac{k_{B}}{e} \right) \left[-y + \delta_{r}(y)\right]
\end{equation}

\begin{equation}
\label{delta}
\delta_{r}(y) = \frac{(r+2) F_{r+1}(y)}{(r+1)F_{r}(y)}
\end{equation}

\noindent where $y$=$(\mu / k_{B}T)$ is the reduced chemical potential, \emph{r} is the scattering parameter for the energy-dependent relaxation time, $\tau (E) = \tau_{0} E^{r}$, which for charged impurities can be taken to be $3/2$,\cite{Ziman,Okuda,Delugas} and $F_{r}$ is the Fermi integral. Using the experimental values of the charge density measured by Hall effect, average $m^{*}=1.5 m_{e}$, from Shubnikov-de Haas \cite{Jalan2010,Caviglia2010} and the following relation to calculate the chemical potential:

\begin{equation}
\label{n}
n = z \left( \frac{2 \pi m^{*} k_{B} T}{h^{2}} \right)^{\frac{3}{2}} \frac{2}{\sqrt{\pi}} F_{\frac{1}{2}}(y)
\end{equation}

the thermoelectric power can be calculated from (1), taking the band degeneracy $z$ as the only adjustable parameter. The results of the calculation are shown in Figure \ref{Imagen3} (c) and (d), along with the experimental data. There is a very good agreement between the calculation and the experimental data. But it is also clear from these plots that films with larger $\Delta$t fit better to the curves with $z=4$, in both STO and LSAT, while the films grown at higher oxygen pressures ($\Delta$t small) fit better to the $z$=6 curve.


\begin{figure}[htbp]
	\centering
		\hspace{-6px}\includegraphics[width=0.50\textwidth]{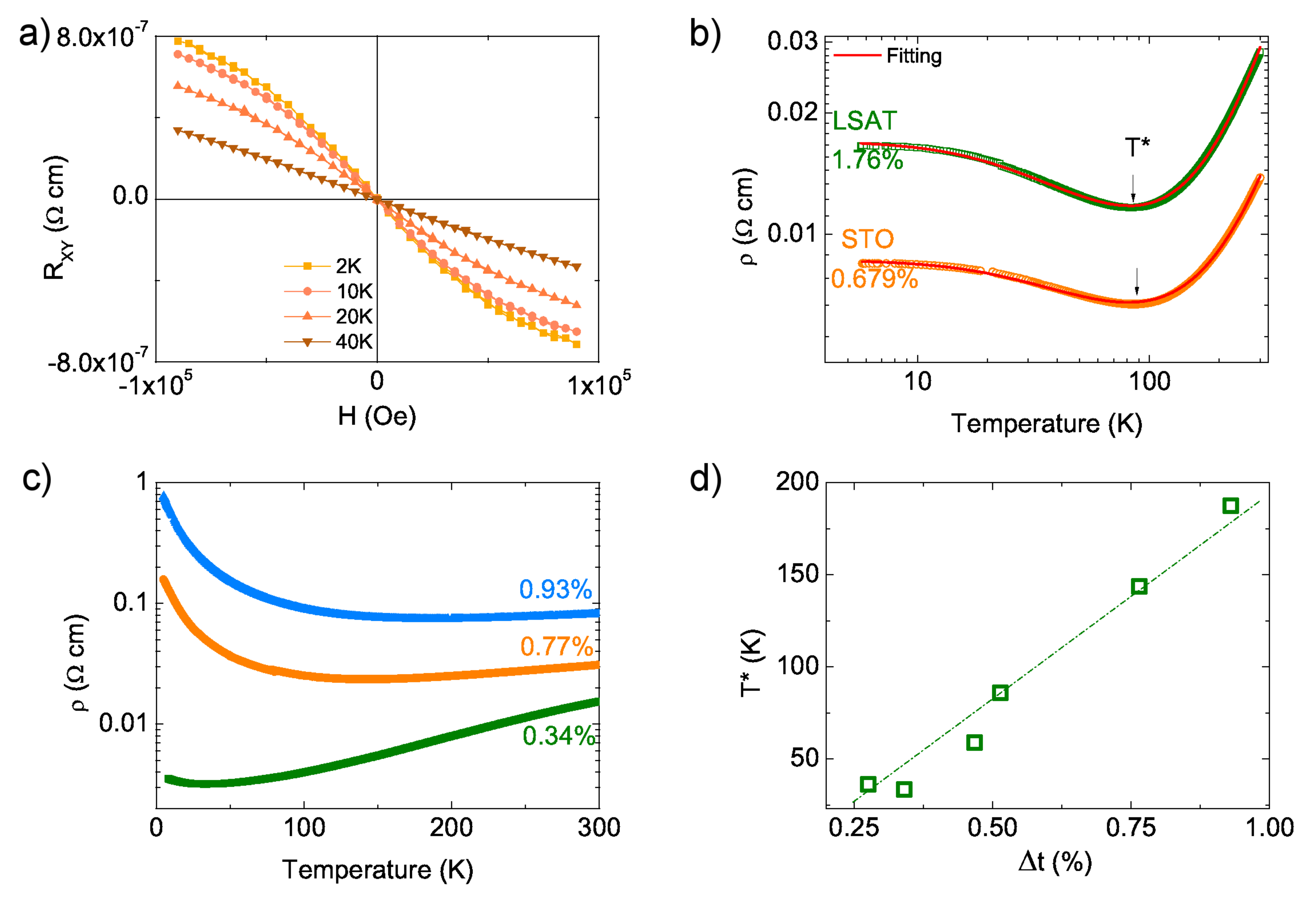}
	\vspace{-20px}\caption{\footnotesize{a) Hall resistivity at different temperatures. The non-linear behavior at high field disappears above T*. b) The resistivity for two samples grown at the same oxygen pressure on different substrates, indicating the T* and the corresponding $\Delta$t. The solid line represents the fitting to the Kondo equation (see supporting information). c) Temperature dependence of the resistivity for Nb:STO thin films deposited on STO substrates at different oxygen pressures (resulting in a different tetragonal distortion). d) Variation of $T^{*}$, as a function of the tetragonal distortion $\Delta$t.}}
	\label{Imagen4}
\end{figure}


Therefore, thermoelectric power and ab-initio calculations support a scenario in which an additional band splitting is produced in Nb:STO due to the presence of V$_{Sr}$, whose amount is extremely sensitive to deposition conditions. Moving the Fermi energy across these bands will change the relative contribution from heavy/light electrons to the transport properties of the system. Actually, a change from one- to two-carrier transport has been suggested at a universal critical density in LAO/STO interfaces.\cite{Joshua2012, Shalom2010} Particularly, due to the contribution of lower mobility carriers at large fields, this results in a non-linear contribution to the Hall effect at low temperatures.

The Hall data is shown in Figure \ref{Imagen4}a for one of our films with $\Delta$t = $0.68\%$. At low temperature there is a clear deviation from the low-field linear behavior, similar to observations in LAO/STO interfaces.\cite{Shalom2010, Joshua2012}

This behavior is observed up to a given temperature T*, which marks a minimum in the temperature dependence of the resistivity.
An upturn and saturation in the low temperature resistivity as shown in Figure \ref{Imagen4}b) was previously reported in electrolyte-gated STO, associated to Kondo effect from local Ti$^{3+}$ magnetic moments.\cite{Lee2011,Parkin2012} However, as shown in Figure \ref{Imagen4}c), T*  increases linearly with $\Delta$t, showing that the change in the conduction mechanism at low temperature is determined by the total concentration of defects. This, along with the absence of any anomaly in the Seebeck coefficient at the Kondo temperature, suggests that in this case the transition to a thermally activated behavior observed below T* is most probably due to conventional vacancy scattering, and not to Kondo effect.


\begin{figure}[htbp]
	\centering
		\hspace{-6px}\includegraphics[width=0.50\textwidth]{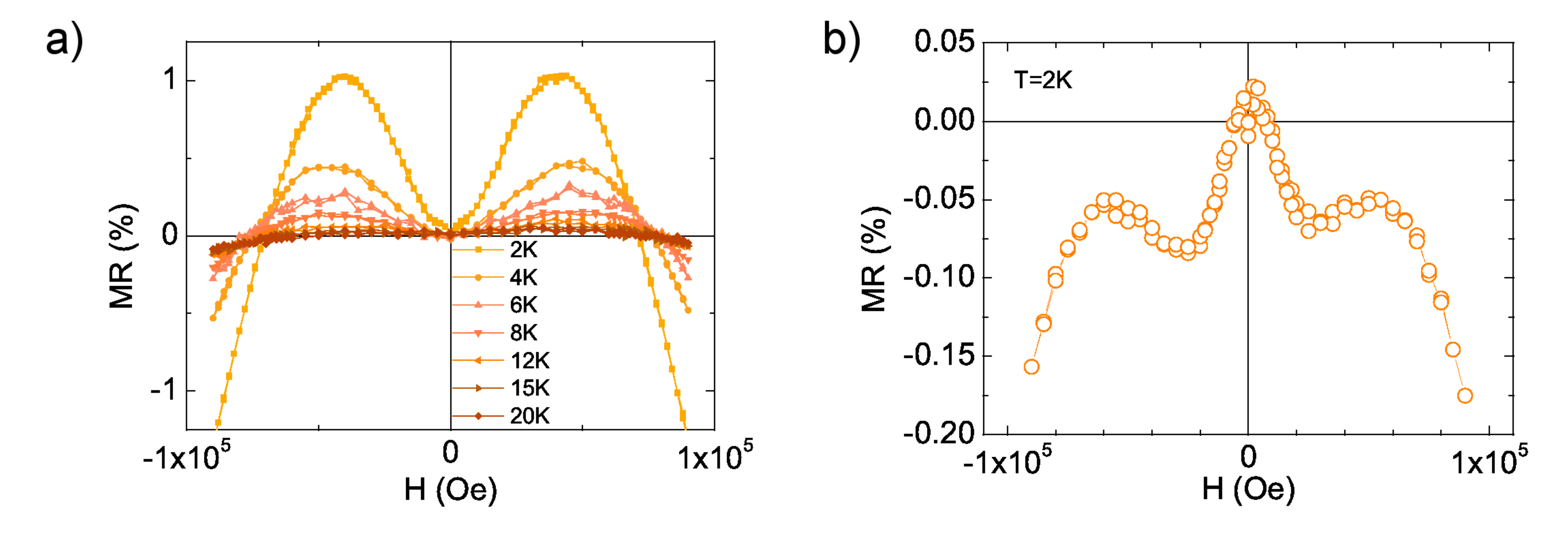}
	\vspace{-20px}\caption{\footnotesize{a) Anisotropic magnetoresistance, for a 50 nm thick film deposited on STO at 10 mTorr ($\Delta$t$\approx$0.68\%). The magnetic field is applied in the plane of the film, perpendicular to the current. b) Magnetoresistance of the same film measured at 2 K, with the magnetic field applied perpendicular to the surface of the film.}}
	\label{Imagen5}
\end{figure}


On the other hand, a non-linear Hall effect could be due to an anomalous contribution from a magnetically ordered phase.\cite{Shalom2010b}
In fact, the presence of localized magnetic moments and magnetic order in doped STO and LAO/STO interfaces is a controversial issue, whose origin is still under discussion. \cite{Brinkman2007,Santander-Syro2014, Seri2009, Shalom2009}

An important test for the presence of localized magnetic moments and spontaneous magnetic ordering will come from the study of the anisotropic magnetoresistance (AMR). Below T*, the transverse MR (H in plane, perpendicular to the current, Fig. 5 a) shows a change of sign at $\approx$ 5 T, while the perpendicular MR (H out of plane, Fig. 5 b) is a non-monotonically-decreasing function of H. This behavior is in principle compatible with a weak anti-localization (WAL) contribution to the resistivity in the presence of a strong spin-orbit coupling.\cite{Caviglia2010} However, we have observed this behavior in 50 nm thick, highly distorted films. Moreover, the anisotropy in the MR (see supporting information), suggests that the source of this spin-orbit scattering is a magnetic phase, rather than independent magnetic moments (isotropic) at Ti$^{3+}$ sites. We want to remark that a very similar behavior was observed in STO quantum wells embedded in antiferromagnetic SmTiO$_3$, ascribed to a magnetic proximity effect, \cite{Stemmer2014}, as well as in LAO/STO interfaces. \cite{Budhani2014}
Also, although Fe impurities are normally associated to Ti, the fact that the MR is observed below a minimum which depends on strain and not only on the composition of the films, makes highly improbable that the AMR effect is related to the presence of magnetic Fe impurities in the films.


In summary, we have demonstrated that even a small concentration of cationic defects in epitaxial thin films of electron-doped STO produce a homogeneous tetragonal distortion of the system. This has a profound impact on the electronic band degeneracy and therefore on the magnetotransport properties, as we have shown by combining thermopower measurements and ab-initio calculations. On the other hand, a non-linear contribution to the Hall effect, and anisotropic MR compatible with a magnetically ordered phase, has been observed in distorted thin films at low temperature. Therefore, our results show that a similar phenomenology to that of LAO/STO interfaces and quantum wells can be achieved in e-doped STO thin films, after proper control of cationic vacancies during growth.

This work was supported by the European Research Council ERC
StG-259082 2DTHERMS, MINECO of Spain MAT2013--44673-R, and
Xunta de Galicia Projects No. 2012-CP071 and EM2013/037. V.P. acknowledges support from the Spanish Government through the Ram\'on y Cajal Program.

Author contribution statement: A.S. and E.F-V. contributed equally to this work.



%

\end{document}